\documentclass[a4paper,12pt]{article}
\usepackage{jcappub}
\usepackage[symbol]{footmisc}
\usepackage{amsmath,amssymb}

\newcommand{\nc}{\newcommand*} 
\newcommand{\mU}{{\mathcal{U}}}

\nc{\Eq}[1]{Eq.~\eqref{#1}}     
\nc{\Fig}[1]{Fig.~\ref{#1}}     
\nc{\Table}[1]{Table~\ref{#1}}  
\nc{\Sec}[1]{Sec.~\ref{#1}}     
\def\({\left(}
\def\){\right)}
\def\[{\left[}
\def\]{\right]}
\def\e{\begin{equation}}
\def\q{\end{equation}}
\def\m{\begin{eqnarray}}
\def\n{\end{eqnarray}}

\begin{document}

\title{Prospects for Taiji to detect a gravitational-wave background from cosmic strings}

\author{Zu-Cheng~Chen,$^{a,b,c,d}$}
\author{Qing-Guo Huang,$^{e,f,g}$}
\author{Chang Liu,$^{h,e,f}$}
\author{Lang~Liu,\note{Corresponding author.}$^{c,d,*}$}
\author{Xiao-Jin Liu,$^{c,d}$}
\author{You~Wu,$^{i,*}$}
\author{Yu-Mei Wu,$^{e,f}$}
\author{Zhu~Yi,$^{d,*}$}
\author{and Zhi-Qiang~You$^{j}$}

\affiliation{$^a$Department of Physics and Synergetic Innovation Center for Quantum Effects and Applications, Hunan Normal University, Changsha, Hunan 410081, China}
\affiliation{$^b$Institute of Interdisciplinary Studies, Hunan Normal University, Changsha, Hunan 410081, China}
\affiliation{$^c$Department of Astronomy, Beijing Normal University, Beijing 100875, China}
\affiliation{$^d$Advanced Institute of Natural Sciences, Beijing Normal University, Zhuhai 519087, China}
\affiliation{$^e$School of Fundamental Physics and Mathematical Sciences, Hangzhou Institute for Advanced Study, UCAS, Hangzhou 310024, China}
\affiliation{$^f$School of Physical Sciences, University of Chinese Academy of Sciences, No. 19A Yuquan Road, Beijing 100049, China}
\affiliation{$^g$CAS Key Laboratory of Theoretical Physics, Institute of Theoretical Physics, Chinese Academy of Sciences, Beijing 100190, China}
\affiliation{$^h$Center for Gravitation and Cosmology, College of Physical Science and Technology, Yangzhou University, Yangzhou, 225009, China}
\affiliation{$^i$College of Mathematics and Physics, Hunan University of Arts and Science, Changde, 415000, China}
\affiliation{$^j$Henan Academy of Sciences, Zhengzhou 450046, Henan, China}

\emailAdd{zuchengchen@hunnu.edu.cn}
\emailAdd{huangqg@itp.ac.cn}
\emailAdd{liuchang@alumni.itp.ac.cn}
\emailAdd{liulang@bnu.edu.cn}	
\emailAdd{xliu.astro@bnu.edu.cn}
\emailAdd{youwuphy@gmail.com}	
\emailAdd{ymwu@ucas.ac.cn} 
\emailAdd{yz@bnu.edu.cn}	
\emailAdd{you\_zhiqiang@whu.edu.cn}

\abstract{
Recently, multiple pulsar timing array collaborations have presented compelling evidence for a stochastic signal at nanohertz frequencies, potentially originating from cosmic strings. Cosmic strings are linear topological defects that can arise during phase transitions in the early Universe or as fundamental strings in superstring theory. This paper focuses on investigating the detection capabilities of Taiji, a planned space-based gravitational wave detector, for the gravitational wave background generated by cosmic strings. By analyzing simulated Taiji data and utilizing comprehensive Bayesian parameter estimation techniques, we demonstrate a significant improvement in precision compared to the NANOGrav 15-year data set, surpassing it by an order of magnitude. This highlights the enhanced measurement capabilities of Taiji. Consequently, Taiji can serve as a valuable complementary tool to pulsar timing arrays in validating and exploring the physics of cosmic strings in the early Universe.
}
	
\maketitle

\section{Introduction}
The direct detection of gravitational waves (GWs) from compact binary coalescences by ground-based detectors~\cite{LIGOScientific:2018mvr,LIGOScientific:2020ibl,LIGOScientific:2021djp} has revolutionized our ability to test gravity theories in the strong-field regime~\cite{LIGOScientific:2019fpa,LIGOScientific:2020tif,LIGOScientific:2021sio} and study the properties of individual systems and the broader population of GW sources \cite{LIGOScientific:2018jsj,Chen:2018rzo,Chen:2019irf,LIGOScientific:2020kqk,Chen:2021nxo,KAGRA:2021duu,Chen:2022fda,Liu:2022iuf,You:2023ouk}. 
Stochastic gravitational wave backgrounds (SGWBs) represent an additional class of gravitational wave sources that yet to be detected. 
Recently, several pulsar timing array (PTA) collaborations, including NANOGrav~\cite{NANOGrav:2023hde,NANOGrav:2023gor}, PPTA~\cite{Zic:2023gta,Reardon:2023gzh}, EPTA~\cite{Antoniadis:2023lym,Antoniadis:2023ott}, and CPTA~\cite{Xu:2023wog}, have independently announced compelling evidence for a stochastic signal in their latest data sets, displaying the Hellings-Downs~\cite{Hellings:1983fr} spatial correlations consistent with an SGWB as predicted by general relativity. 
While the PTA window encompasses a wide range of potential sources~\cite{Li:2019vlb,Vagnozzi:2020gtf,Chen:2021wdo,Wu:2021kmd,Chen:2021ncc,Benetti:2021uea,Ashoorioon:2022raz,PPTA:2022eul,Wu:2023pbt,IPTA:2023ero,Wu:2023dnp,Madge:2023cak,Wu:2023rib,Bi:2023ewq,Chen:2023uiz}, the nature of the observed signal, whether it originates from astrophysical phenomena or cosmological processes, remains the subject of ongoing investigation~\cite{NANOGrav:2023hvm,Antoniadis:2023xlr,King:2023cgv,Niu:2023bsr,Ben-Dayan:2023lwd,Vagnozzi:2023lwo,Fu:2023aab,InternationalPulsarTimingArray:2023mzf,Basilakos:2023jvp}.
The PTA signal can potentially be attributed to various sources~\cite{Bian:2023dnv,Wu:2023hsa,Ellis:2023oxs,Figueroa:2023zhu}, such as the SGWB produced by supermassive black hole binaries~\cite{NANOGrav:2023hfp,Ellis:2023dgf,Shen:2023pan,Bi:2023tib,Barausse:2023yrx}, scalar-induced GWs~\cite{Cai:2019bmk,Yuan:2019udt,Yuan:2019wwo,Chen:2019xse,Yuan:2019fwv,Liu:2021jnw,Dandoy:2023jot,Franciolini:2023pbf,Wang:2023ost,Jin:2023wri,Liu:2023pau,Yi:2023npi,Zhao:2023joc,Harigaya:2023pmw,Balaji:2023ehk,Firouzjahi:2023lzg,Cai:2023dls,Liu:2023ymk,Yi:2021lxc,Yi:2022ymw,Yi:2023tdk,You:2023rmn,Yi:2023mbm,Liu:2023hpw} associated with the formation of primordial black holes~\cite{Liu:2018ess,Chen:2018czv,Liu:2019rnx,Liu:2020cds,Wu:2020drm,Chen:2022qvg,Meng:2022low,Lu:2019sti,Gao:2020tsa,Yi:2020cut,Yi:2020kmq,Zheng:2022wqo,Yi:2022anu,Bhaumik:2023wmw,Bousder:2023ida,HosseiniMansoori:2023mqh,Gouttenoire:2023nzr,Huang:2023chx,Depta:2023qst}, phase transitions~\cite{Addazi:2023jvg,Athron:2023mer,Zu:2023olm,Jiang:2023qbm,Xiao:2023dbb,Abe:2023yrw,Gouttenoire:2023bqy,An:2023jxf,Chen:2023bms}, domain walls~\cite{Kitajima:2023cek,Blasi:2023sej,Babichev:2023pbf}, and cosmic strings~\cite{Chen:2022azo,Kitajima:2023vre,Ellis:2023tsl,Wang:2023len,Antusch:2023zjk,Ahmed:2023pjl,Ahmed:2023rky,Basilakos:2023xof}.

Cosmic strings are linear topological defects that can originate from symmetry-breaking phase transitions at high energies in the early Universe~\cite{Kibble:1976sj,Vilenkin:1981bx,Vilenkin:1984ib}. Alternatively, they can be the fundamental strings of superstring theory or one-dimensional D-branes~\cite{Dvali:2003zj,Copeland:2003bj}, which can extend to astrophysical scales. Once cosmic strings are formed, their intersections can result in reconnections and the formation of loops. These loops undergo relativistic oscillations and eventually decay, emitting GWs in the process. 
PTA observations can potentially detect a cosmic string network through two distinctive signatures. The first signature involves the detection of GW bursts emitted at cusps, see~\textit{e.g.}~\cite{Yonemaru:2020bmr}. The second signature involves observing the SGWB, which arises from the cumulative radiation emitted by all cosmic string loops present throughout cosmic history, see~\textit{e.g.}~\cite{Chen:2022azo,Bian:2022tju}.

The GW energy spectrum associated with cosmic strings spans a wide range of frequencies. If the observed signal in PTAs is indeed attributed to cosmic strings, it is widely anticipated~\cite{Boileau:2021gbr,Wang:2022pav,Wang:2023ltz} that this signal will also be detected by upcoming space-based GW detectors such as LISA~\cite{LISA:2017pwj}, Taiji~\cite{Hu:2017mde,Ruan:2018tsw}, and TianQin~\cite{TianQin:2015yph}, as shown in \Fig{ogw_cs}. Taiji~\cite{Hu:2017mde,Ruan:2018tsw} is a planned space-based GW detection mission scheduled for launch in the 2030s. It consists of three spacecraft, each following a heliocentric orbit, forming a triangle with a length of approximately $3$ million kilometers. With its sensitivity in the range of $10^{-5}\sim 10^{-1}$ Hz, Taiji offers a complementary probe of cosmic strings compared to PTAs. Thus, space-based GW detectors like Taiji play a crucial role in verifying the stochastic signal detected by PTAs and serve as invaluable tools for further validation of cosmic string signatures.

In this paper, we assume that the signal detected by PTAs originates from the SGWB produced by cosmic strings and determine the model parameters using the NANOGrav 15-year data set. By analyzing the simulated Taiji data and applying parameter estimation techniques, we aim to quantify the uncertainties in determining the model parameters. This analysis will provide valuable insights into the precision and accuracy of parameter estimation for the SGWB produced by cosmic strings, as well as the detection capabilities of Taiji. The rest of the paper is structured as follows. In \Sec{CS}, we provide a brief review of the SGWB produced by the cosmic strings and constrain the model parameters with the NANOGrav 15-year data set. 
In \Sec{model}, we introduce the various components of the model that will be used in the simulation and parameter estimation. These components include the Taiji noise model, the foreground from double white dwarfs (DWDs), and the extragalactic compact binary (ECB) foreground.
In \Sec{method}, we describe the methodology for simulating Taiji data, inferring model parameters, and present the results on constraining the model parameters with Taiji, providing insights into Taiji's ability to detect the SGWB produced by cosmic strings.
Finally, we provide the concluding remarks in \Sec{conclusion}. 

\section{\label{CS}Cosmic strings and the constraints by NANOGrav}

\begin{figure}[tbp!]
	\centering
	\includegraphics[width=0.8\linewidth]{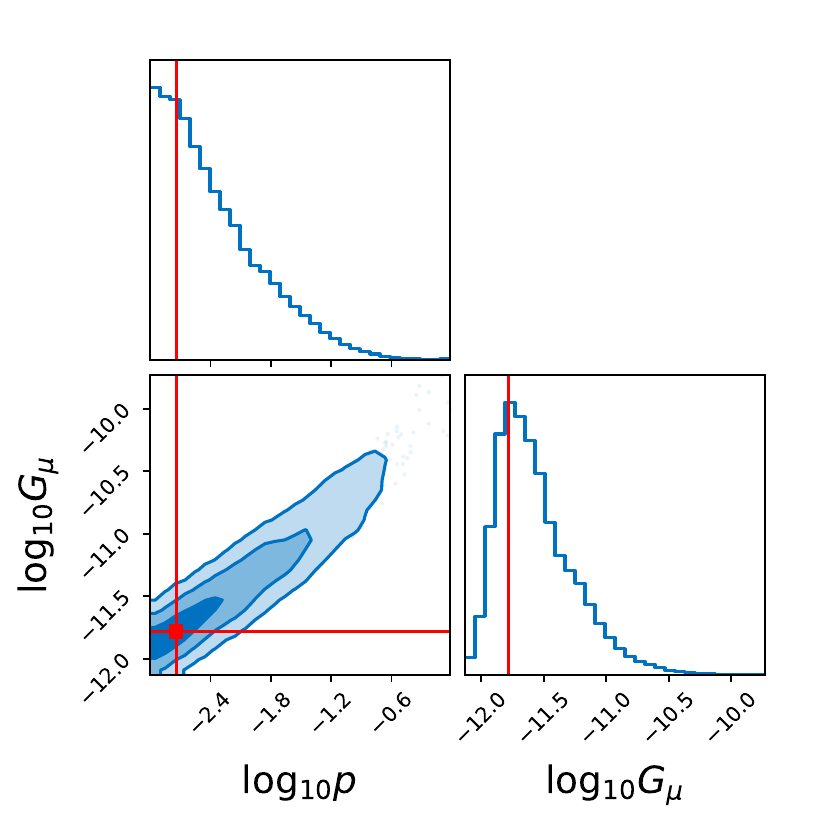}
	\caption{\label{posts_ng}Posteriors for the model parameters obtained from the combined constraints of the NANOGrav 15-year data set and the LVK O3 data. The red point represents the maximum-likelihood values injected into the Taiji data. The plot displays the $1\sigma$, $2\sigma$, and $3\sigma$ confidence contours in the two-dimensional parameter space.}
\end{figure}

In this section, we give a brief review of the SGWB produced by cosmic strings closely following~\cite{Blanco-Pillado:2017oxo}.
A cosmic string network comprises both long (or ``infinite") strings that exceed the size of the cosmic horizon and loops formed by smaller strings.  
When two cosmic strings intersect, there is a possibility for them to undergo a reconnection process with a probability denoted as $p$. This reconnection event results in the formation of loops.
Once loops are formed in the cosmic string network, they undergo relativistic oscillations~\cite{Vilenkin:1981bx}, causing them to emit GWs. The emission of GWs is responsible for the gradual decay of the loops, leading to a reduction in their size over time. 
A cosmic string network grows along with the cosmic expansion and evolves toward the scaling regime in which all the fundamental properties of the network grow proportionally with the cosmic time. 
The scaling regime can be achieved through the formation and subsequent decay of loops. 
The GW spectrum generated by a cosmic string network exhibits an exceptionally wide frequency range, spanning from frequencies below $10^{-16}$ Hz to frequencies exceeding $10^9$ Hz, depending on the sizes of the generated loops.

\begin{figure}[tbp!]
	\centering
	\includegraphics[width=\linewidth]{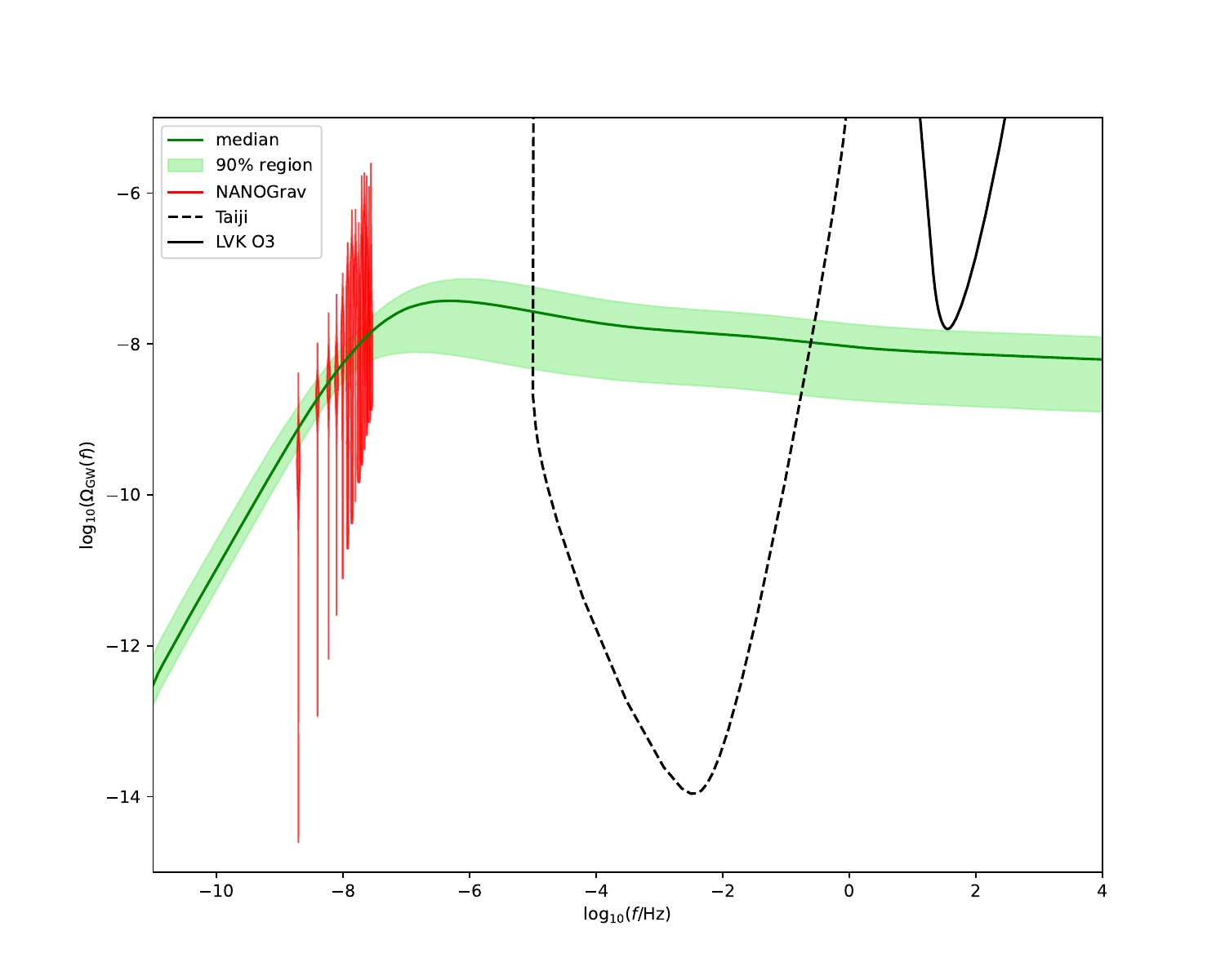}
	\caption{\label{ogw_cs} The posterior predictive distribution of the energy density spectrum of cosmic strings. We also show the free spectrum obtained from the NANOGrav 15-year data set (red color), the power-law integrated sensitivity curves for Taiji (black dashed line) and LVK O3 (black solid line). }
\end{figure}

In this work, we characterize the GW energy spectrum of cosmic strings using two key parameters: the dimensionless tension, $G\mu$, and the reconnection probability, $p$. While classical strings typically have a reconnection probability of $p=1$, in string-theory-inspired models~\cite{Hanany:2005bc} or pure Yang-Mills theory~\cite{Yamada:2022imq,Yamada:2022aax}, it can be less than $1$. 
The dimensionless GW energy density parameter per logarithm frequency, as the fraction of the critical energy density, is~\cite{Blanco-Pillado:2017oxo}
\e
\Omega_{\mathrm{GW}}(f)=\frac{8 \pi G f}{3 H_{0}^{2} p} \rho_{\mathrm{GW}}(t_0, f),
\q
where $G$ is the gravitational constant, $t_0$ is cosmic time today, $H_0=67.4\, \mathrm{km}/\mathrm{s}/\mathrm{Mpc}$ is the Hubble constant taken from Planck 2018~\cite{Planck:2018vyg}, and $\rho_{\mathrm{GW}}$ is the GW energy density per unit frequency given by
\e
\rho_{\mathrm{GW}}(t, f)=G \mu^{2} \sum_{n=1}^{\infty} C_{n} P_{n},
\q
with
\e
C_{n}(f)=\int_{0}^{t_{0}} \frac{d t}{(1+z)^{5}} \frac{2 n}{f^{2}} \mathrm{n}(l, t).
\q
Here, $P_{n}$ is the radiation power spectrum of each individual loop, and $\mathrm{n}(l, t)$ is the density of loops per unit volume per unit range of loop length $l$ at a given cosmic time $t$. 
The SGWB from a network of cosmic strings was computed using a comprehensive end-to-end method in~\cite{Blanco-Pillado:2017oxo}.
The process involved simulating the long string network to extract representative loop shapes, estimating loop shape deformations due to gravitational backreaction using a smoothing model, and computing the GW spectrum for each loop. The distribution of loops over redshift was evaluated by integrating over cosmological time, and the overall emission spectrum was obtained by integrating the GW spectrum of each loop over the redshift-dependent loop distribution. Finally, the current SGWB was obtained by integrating the overall emission spectrum over cosmological time. The simulations covered a wide parameter range of $G\mu \in [10^{-25}, 10^{-8}$], and $f \in [10^{-15}, 10^{10}]$~Hz.
The resulting energy density spectra were made publicly available\footnote{\url{http://cosmos.phy.tufts.edu/cosmic-string-spectra/}}. We refer to~\cite{Blanco-Pillado:2017oxo} for more detailed information regarding the computations and methodology.

Cosmic strings can be a potential explanation for the recently detected PTA signal. To constrain the model parameters for cosmic strings using the NANOGrav 15-year data set, we perform a Bayesian inference following \cite{Bi:2023tib,Liu:2023ymk,Wu:2023hsa,Jin:2023wri,Liu:2023pau} by taking into account the Hellings-Downs correlations. We obtain $\log_{10} p = -2.49^{+1.09}_{-0.47}$ and $\log_{10} G\mu = -11.62^{+0.65}_{-0.29}$, and the resulting posteriors for $G\mu$ and $p$ are shown in \Fig{posts_ng}. 
Furthermore, it is important to note that the cosmic string model employed in this study has been previously constrained in~\cite{NANOGrav:2023hvm} and is referred to as \texttt{SUPER} model therein. 
The posterior distributions for $G\mu$ and $p$ derived in our analysis, as illustrated in \Fig{posts_ng}, are broadly consistent with those reported in~\cite{NANOGrav:2023hvm}, specifically shown in Fig.~11 therein.
It is also worth noting that the analysis incorporates the constraint that the GW energy density from cosmic strings should not exceed the upper limits derived by the LIGO-Virgo-KAGRA (LVK) collaboration, as null detection of the SGWB has been obtained by LVK O3 data.
The posterior predictive distribution for the GW energy density is illustrated in \Fig{ogw_cs}, along with the power-law integrated sensitivity curves of Taiji and LVK O3. 
The results indicate that the wide frequency range of the energy density makes it detectable by the Taiji detector. 
In what follows, we will focus on examining the detection capability of the Taiji detector for the SGWB generated by cosmic strings.


\section{\label{model}Model components}

\subsection{Taiji noise model}
With the triangular geometry of the Taiji detector, the interferometric phase differences can be combined using the time delay interferometry (TDI) techniques to mitigate laser frequency noise~\cite{Tinto:2001ii,Tinto:2002de}.
This results in the creation of three distinct GW measurement channels known as the $X, Y, Z$ TDI channels~\cite{Vallisneri:2012np}. 
The precise determination of the noise properties in the Taiji mission is a significant technical challenge. The full complexity of the noise properties is extensive, and it is beyond the scope of this discussion to delve into all the details. 
Here, we assume that the SGWB signal in the $X, Y, Z$ channels is stationary and uncorrelated with the Taiji instrument noise. 
Additionally, we simplify the instrument noise by assuming it consists of two components: test mass acceleration noise and optical path length fluctuation noise. 
We also assume that these instrumental noise sources are identical in each spacecraft and that the arm lengths of the Taiji instruments are equal, forming an equilateral triangle with the arm lengths $L_1=L_2=L_3=L=3 \times 10^9\, \mathrm{m}$.
Under these assumptions, the cross-spectra and response functions of the $X, Y, Z$ channel combinations are identical~\cite{Flauger:2020qyi}.

The noise components in the Taiji instrument can be encapsulated by two effective functions.
The first function represents the high-frequency component originating from the optical metrology system and is characterized by the power spectral density (PSD) as (see \textit{e.g.}~\cite{Ren:2023yec})
\e 
P_{\mathrm{oms}}(f) = P^2 \times 10^{-24} \frac{1}{\mathrm{Hz}}\left[1+\left(\frac{2\,\mathrm{mHz}}{f}\right)^4\right]\left(\frac{2 \pi f}{c} \mathrm{m}\right)^2,
\q 
where $f$ is the frequency, $c$ is the speed of light, and $P=8$~\cite{Luo:2019zal}. The second function captures the low-frequency component associated with test mass acceleration noise and is described by the PSD as (see \textit{e.g.}~\cite{Ren:2023yec})
\begin{equation}
P_{\mathrm{acc}}(f) = A^2 \times 10^{-30} \frac{1}{\mathrm{Hz}}\left[1+\left(\frac{0.4\, \mathrm{mHz}}{f}\right)^2\right]\left[1+\left(\frac{f}{8\, \mathrm{mHz}}\right)^4\right]\left(\frac{1}{2 \pi f c} \frac{\mathrm{m}}{\mathrm{s}^2} \right)^2,
\end{equation}
where $A=3$~\cite{Luo:2019zal}.
The total PSD for the noise auto-correlation is
\begin{equation}\label{auto}
N_{a a}(f, A, P)=16 \sin ^2\left(\frac{f}{f_*}\right)\left\{\left[3+\cos \left(\frac{2f}{f_*}\right)\right] P_{\mathrm{acc}}(f, A)+P_{\mathrm{oms}}(f, P)\right\},
\end{equation}
where $f_*=c / (2 \pi L)$ is a characteristic frequency determined by the arm length $L$. 
The noise cross-spectra are
\begin{equation}\label{cross}
N_{a b}(f, A, P)=-8 \sin ^2\left(\frac{f}{f_*}\right) \cos \left(\frac{f}{f_*}\right)\left[4 P_{\mathrm{acc}}(f, A)+P_{\mathrm{oms}}(f, P)\right],
\end{equation}
where $a, b \in\{\mathrm{X}, \mathrm{Y}, \mathrm{Z}\}$ and $a \neq b$. It is worth noting that, under these assumptions, the noise covariance matrix is real.

For convenience, linear combinations of the channels can be used. In the case of the Taiji instrument, three specific channels are commonly chosen: the ``noise orthogonal" channels $A$ and $E$, and the ``null" channel $T$. These channels are defined as follows:
\begin{equation}
\left\{\begin{array}{l}
A=\frac{1}{\sqrt{2}}(Z-X), \\
E=\frac{1}{\sqrt{6}}(X-2 Y+Z), \\
T=\frac{1}{\sqrt{3}}(X+Y+Z) .
\end{array}\right.
\end{equation}
These combinations are chosen to achieve noise orthogonality and a reduced sensitivity to GWs in the null channel $T$. Under the AET basis, the noise spectra can be expressed as
\m 
N_{\mathrm{A}} &=&N_{\mathrm{E}}=N_{\mathrm{XX}}-N_{\mathrm{XY}},\\
N_{\mathrm{T}} &=&N_{\mathrm{XX}}+2 N_{\mathrm{XY}}.
\n
By substituting the noise auto- and cross-spectra from \Eq{auto} and \Eq{cross} into the above expressions, we obtain
\begin{equation}
N_{\mathrm{A,E}}=8 \sin ^2\left(\frac{f}{f_*}\right)  \left\{4\left[1+\cos \left(\frac{f}{f_*}\right)+\cos^2\left(\frac{f}{f_*}\right)\right] P_{\mathrm{acc}} +\left[2+\cos \left(\frac{f}{f_*}\right)\right] P_{\mathrm{oms}}\right\},
\end{equation}
and
\begin{equation}
N_{\mathrm{T}}=16 \sin ^2\left(\frac{f}{f_*}\right)  \left\{2\left[1-\cos \left(\frac{f}{f_*}\right)\right]^2 P_{\mathrm{acc}} + \left[1-\cos \left(\frac{f}{f_*}\right)\right] P_{\mathrm{oms}} \right\}. 
\end{equation}

For convenience, we introduce the equivalent energy spectral density
\begin{equation}
\Omega_{\alpha}(f)= S_{\alpha}(f) \frac{4 \pi^2 f^3}{3 H_0^2},
\end{equation}
where the indices $\alpha \in \{\mathrm{A, E, T}\}$ run over the different channel combinations, and $S_{\alpha}$ represents the noise spectral densities for the $\alpha$ channel. The noise spectral densities are defined by
\m
S_\mathrm{A}(f)&=&S_\mathrm{E}(f)=\frac{N_\mathrm{A}(f)}{\mathcal{R}_\mathrm{A, E}(f)},\\
S_\mathrm{T}(f)&=&\frac{N_\mathrm{T}(f)}{\mathcal{R}_\mathrm{T}(f)},
\n
where $\mathcal{R}_\alpha$ are the response functions for the $\alpha$ channel.
There exist approximate expressions for the response functions given in~\cite{Smith:2019wny}
\m
\mathcal{R}_\mathrm{A}^{\mathrm{Fit}}(f)&=&\mathcal{R}_\mathrm{E}^{\mathrm{Fit}}(f)=\frac{9}{20}|W(f)|^2\left[1+\left(\frac{f}{4 f_* / 3}\right)^2\right]^{-1}, \\
\mathcal{R}_\mathrm{T}^{\mathrm{Fit}} &\simeq& \frac{1}{4032}\left(\frac{f}{f_*}\right)^6|W(f)|^2\left[1+\frac{5}{16128}\left(\frac{f}{f_*}\right)^8\right]^{-1},
\n
where $W(f)=1-e^{-2 i f / f_*}$.
However, in this work, we use the analytical expressions derived in~\cite{Wang:2021owg} for the response functions. 

\subsection{Double white dwarf foreground}

According to the astrophysical population models~\cite{Korol:2020lpq,Korol:2021pun}, it is estimated that there are tens of millions of double white dwarf (DWD) binaries in our Milky Way galaxy. These DWD binaries may simultaneously emit GWs within the frequency range from $10^{-5}\, \mathrm{Hz}$ to $0.1\, \mathrm{Hz}$~\cite{Karnesis:2021tsh}. 
Only a small number of these DWD binaries are resolvable, and the majority of these sources remain unresolved and collectively contribute to a stochastic ``galactic foreground" or ``confusion noise" for Taiji~\cite{Liu:2023qap}. 
The DWD foreground can be fitted by a polynomial function $S_c(f)$ in the logarithmic scale as~\cite{Liu:2023qap}
\begin{equation}
S_{\mathrm{DWD}}(f)=\exp \left(\sum_{i=0}^5 a_i\left(\log \left(\frac{f}{\mathrm{mHz}}\right)\right)^i\right) \mathrm{Hz}^{-1} .
\end{equation}
The fitting is applicable for the frequency range $0.1\, \mathrm{mHz}<f<10\, \mathrm{mHz}$. The values of parameters $a_i$ for a $4$-year observation are $a_0 =-85.5448$, $a_1=-3.23671$, $a_2=-1.64187$, $a_3=-1.14711$, $a_4=0.0325887$, $a_5=0.187854$. Then the corresponding dimensionless energy spectral density is
\begin{equation}
\Omega_{\mathrm{DWD}}(f)=S_{\mathrm{DWD}}(f) \frac{4 \pi^2 f^3}{3 H_0^2}.
\end{equation} 

In this study, a broken power law is utilized to approximate the dimensionless energy spectral density $\Omega_{\mathrm{DWD}}$ of the DWD foreground. The expression for $\Omega_{\mathrm{DWD}}(f)$ is given by
\begin{equation}
\Omega_{\mathrm{DWD}}(f)=\frac{A_1\left(f / f_*\right)^{\alpha_1}}{1+A_2\left(f / f_*\right)^{\alpha_2}},
\end{equation}
where the parameters are set as follows: $A_1=3.98\times 10^{-16}$, $A_2 = 4.79\times 10^{-7}$, $\alpha_1= -5.7$, and $\alpha_2=-6.2$.
The spectral shape of the DWD foreground is characterized by a broken power law due to the physical limitation imposed by the respective radii of the two white dwarfs in each binary. At high frequencies, the number of DWDs decreases, resulting in a change in the spectral behaviour.

\subsection{Extragalactic compact binary foreground}

The presence of a background arising from compact binaries consisting of black holes and neutron stars in other galaxies is anticipated~\cite{Chen:2018rzo}. However, the extragalactic compact binary (ECB) foreground has not yet been detected by ground-based GW observatories.
In our model, we approximate the energy spectral density of the ECB foreground using a power law:
\begin{equation}
\Omega_{\mathrm{ECB}}(f) = A_{\mathrm{ECB}}\left(\frac{f}{f_{\mathrm{ref}}}\right)^{\alpha_{\mathrm{ECB}}},
\end{equation}
where $f_{\mathrm{ref}} = 25 \mathrm{Hz}$ represents the reference frequency, and $\alpha_{\mathrm{ECB}} = \frac{2}{3}$ denotes the power-law index. The amplitude $A_{\mathrm{ECB}}$ is estimated to be $1.8 \times 10^{-9}$ based on the analysis presented in~\cite{Chen:2018rzo}.

\section{\label{method}Methodology and Results}


\begin{figure}[htbp!]
	\centering
	\includegraphics[width=\linewidth]{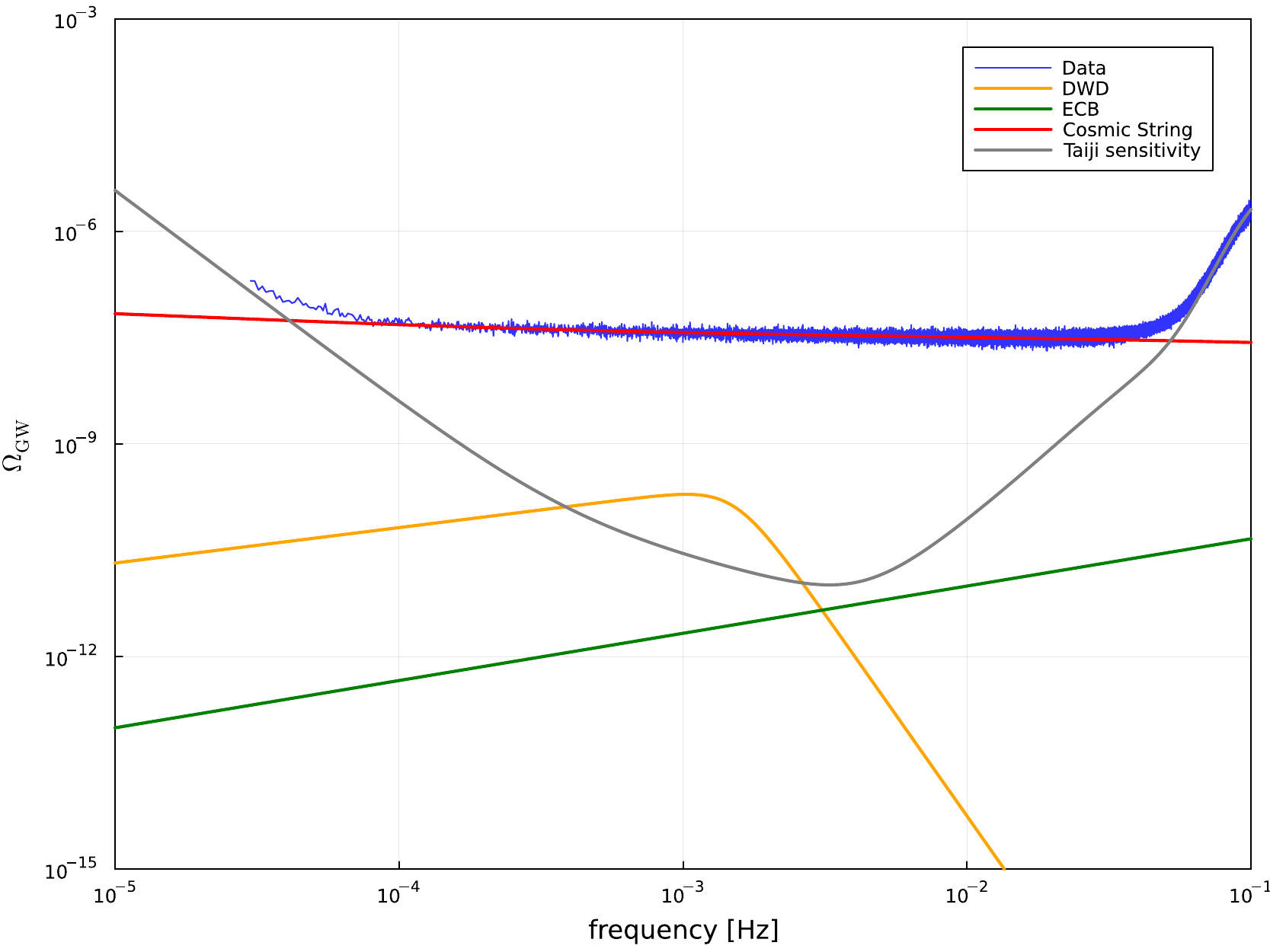}
	\caption{\label{data}The simulated A-channel data (blue) for Taiji in the frequency domain. The DWD foreground (orange), ECB foreground (green), and SGWB produced by cosmic strings (red) are also shown. Additionally, the Taiji sensitivity curve is displayed in terms of the dimensionless energy spectral density $\Omega_{\mathrm{GW}}(f)$.
 }
\end{figure}
We simulate the Taiji data closely following the procedures presented in Refs.~\cite{Caprini:2019pxz,Flauger:2020qyi}. We now briefly outline the methodology. 

In our simulations, we consider a scenario where the data is collected over a full mission duration of $4$ years, with an estimated efficiency of $75\%$, resulting in an effective observation time of $3$ years. 
The TDI data is divided into approximately $N_\mathrm{c} = 94$ chunks (or data segments) of $11.5$ days each, following~\cite{Caprini:2019pxz,Flauger:2020qyi}. 
The frequency range of interest for the Taiji mission extends from $3 \times 10^{-5}\, \mathrm{Hz}$ to $5 \times 10^{-1}\, \mathrm{Hz}$. This corresponds to an average of around $5 \times 10^5$ data points per frequency, with a frequency resolution of approximately $10^{-6}\, \mathrm{Hz}$. In total, the simulation involves roughly $5 \times 10^7$ data points.

To analyze the data stream, we begin by performing a Fourier transform. 
Given that the time stream is real, we will proceed by considering only positive frequencies. We can express the time data stream as
\begin{equation}
d(t) = \sum_{f=f_{\min}}^{f_{\max}} \left[d(f) e^{-2 \pi i f t} + d^*(f) e^{2 \pi i f t}\right].
\end{equation}
Here, $d(f)$ represents the Fourier amplitudes at frequency $f$, and $d^*(f)$ denotes the complex conjugate of $d(f)$. 
We assume that the SGWB and the noise are both stationary so that $\left\langle d(t) d\left(t^{\prime}\right)\right\rangle=f(t-t^{\prime})$ and $\left\langle d(t) \right\rangle = 0$. 
The ensemble averages of the Fourier coefficients satisfy
\begin{equation}
\left\langle d(f) d\left(f^{\prime}\right)\right\rangle=0 \quad \text { and } \quad\left\langle d(f) d^*\left(f^{\prime}\right)\right\rangle=D(f) \delta_{f f^{\prime}},
\end{equation}
indicating that the Fourier coefficients at different frequencies are uncorrelated, and the Fourier coefficients at the same frequency are correlated with the correlation strength determined by the PSD, $D(f)$.

\begin{table}
\centering
\begin{tabular}{c|c|c|c}
\hline\hline
Parameter & Prior & Injected value  & Recovered value \\ 
\hline
$A$ & $\mU(2.95, 3.05)$ & $3$  & $2.996^{+0.013}_{-0.013}$ \\ 
$P$ & $\mU(7.99, 8.01)$ & $8$  & $8.000^{+0.002}_{-0.002}$ \\ 
$\log_{10} A_1$ & $\mU(-16, -15)$ & $-15.4$  & $-15.49^{+0.34}_{-0.36}$ \\ 
$\alpha_1$ & $\mU(-6, -5.5)$ & $-5.7$  & $-5.78^{+0.24}_{-0.20}$ \\ 
$\log_{10} A_2$ & $\mU(-6.5, -6)$  & $-6.32$  & $-6.35^{+0.29}_{-0.13}$ \\ 
$\alpha_2$ & $\mU(-6.5, -6)$ & $-6.2$  &$-6.12^{+0.11}_{-0.27}$ \\ 
$\log_{10} A_{\mathrm{ECB}}$ & $\mU(-9, -8.5)$ & $-8.74$  & $-8.66^{+0.14}_{-0.27}$ \\ 
$\alpha_{\mathrm{ECB}}$ & $\mU(0.5, 1)$ & $2/3$  & $0.58^{+0.19}_{-0.08}$ \\ 
$\log_{10} p$ & $\mU(-2.78, -2.7)$ & $-2.74 $  & $-2.73^{+0.016}_{-0.013}$ \\ 
$\log_{10} G\mu$ & $\mU(-11.9, -11.65)$ & $-11.78$  & $-11.77^{+0.04}_{-0.03}$ \\ 
\hline
\end{tabular}
\caption{\label{tab:priors}Model parameters and their priors used in the Bayesian inferences. The injected values and the recovered values are also presented, with the median and $90\%$ equal-tail uncertainties quoted. Here, $\mU$ denotes the uniform distribution.}
\end{table}

To generate a simulated signal, we utilize a Gaussian distribution with variance $D(f) / 2$ to generate independent random variables for the real and imaginary parts of the Fourier coefficients. By assuming the signal and noise to be Gaussian processes, their statistical properties are fully characterized by the power spectra $\Omega_{\mathrm{GW}}(f)$ and $\Omega_{\mathrm{A,E,T}}(f)$.
Specifically, at each frequency, we generate the quantity
\m
S_i & =&\left|\frac{G_{i 1}\left(0, \sqrt{\Omega_{\mathrm{GW}}\left(f_i\right)}\right)+i G_{i 2}\left(0, \sqrt{\Omega_{\mathrm{GW}}\left(f_i\right)}\right)}{\sqrt{2}}\right|^2, \\
N_i & =&\left|\frac{G_{i 3}\left(0, \sqrt{\Omega_{\mathrm{A,E,T}}\left(f_i\right)}\right)+i G_{i 4}\left(0, \sqrt{\Omega_{\mathrm{A,E,T}}\left(f_i\right)}\right)}{\sqrt{2}}\right|^2 .
\n
In the above expression $G_{i 1}(M, \sigma), \ldots, G_{i 4}(M, \sigma)$ are four real numbers randomly drawn from a Gaussian distribution of average $M$ and variance $\sigma$. These four numbers represent the real and imaginary parts of the Fourier coefficients of both the signal and the noise.
To generate the data, we combine the values of the signal power $S_i$ and noise power $N_i$ by adding them together, assuming that the noise is uncorrelated with the signal. Mathematically, we have
\begin{equation}
D_i=S_i+N_i.
\end{equation}
For each frequency $f_i$, we generate $N_\mathrm{c}$ data values denoted as $\{D_{i1}, D_{i2}, \ldots, D_{iN_\mathrm{c}}\}$. Subsequently, we calculate their average $\bar{D}_i$. In \Fig{data}, we provide an illustration of the simulated data. The parameter values for simulation are summarized in \Table{tab:priors}.

\begin{figure}[tbp!]
	\centering
	\includegraphics[width=\linewidth]{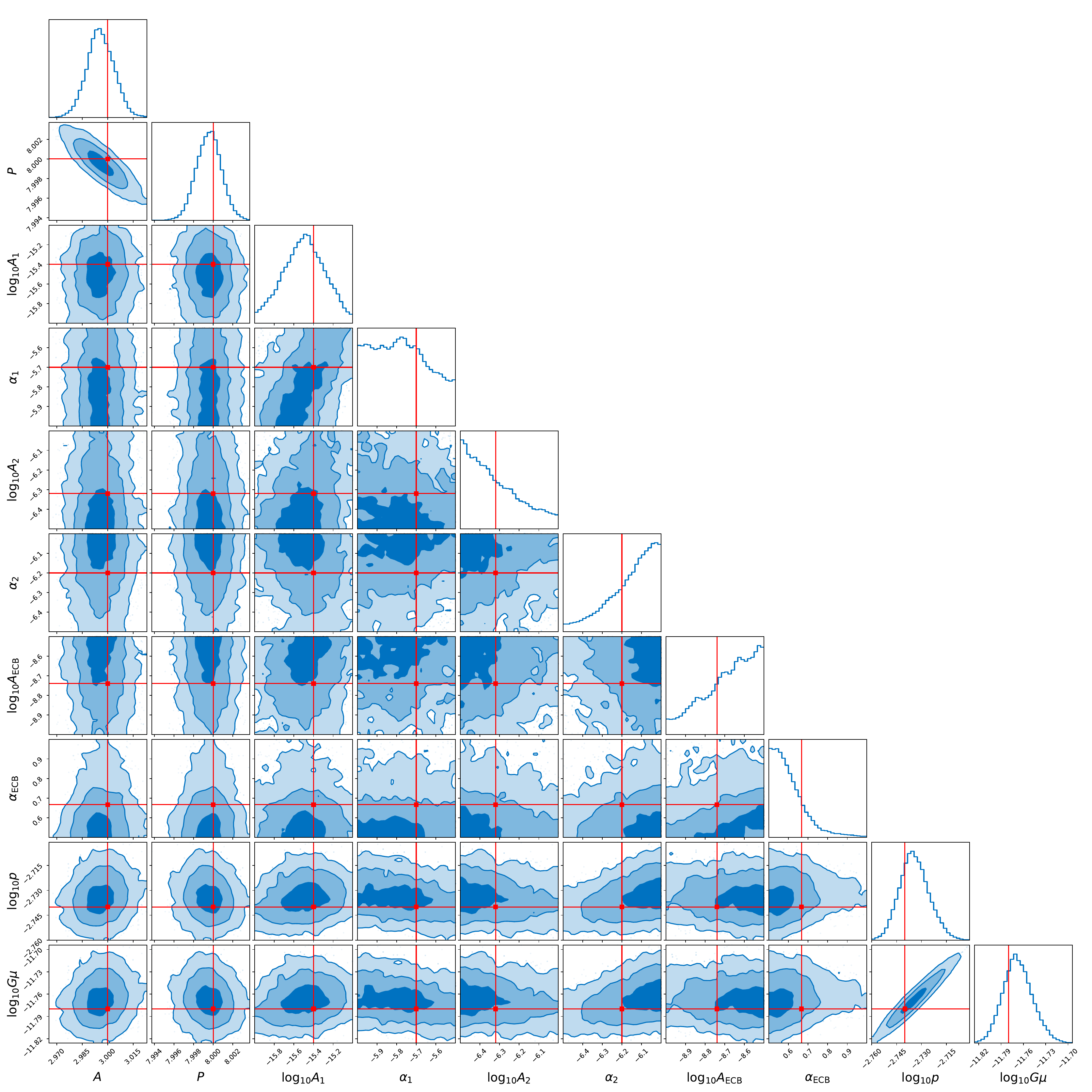}
	\caption{\label{posts_Taiji}Posteriors for the model parameters inferred by the Taiji detector. The injected values for each parameter are highlighted in red. The plot depicts the confidence contours corresponding to the $1\sigma$, $2\sigma$, and $3\sigma$ confidence levels in the two-dimensional parameter space.}
\end{figure}

\begin{figure}[tbp!]
	\centering
	\includegraphics[width=0.8\linewidth]{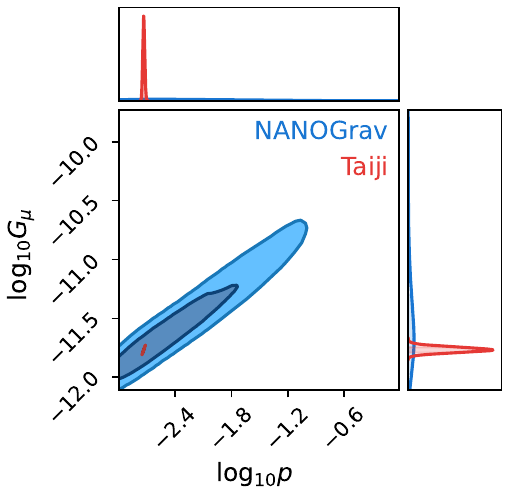}
	\caption{\label{compare}The posterior for the parameters from the cosmic string model, $p$ and $G\mu$, obtained from the NANOGrav 15-year data set (blue) and the Taiji mission (red), respectively.}
\end{figure}
With a linear spacing of $\Delta f=10^{-6} \mathrm{~Hz}$, a significant amount of data points is available at the highest frequencies. To alleviate the computational burden, we adopt a rebinning approach, or ``coarse graining" specifically for frequencies ranging from $f=10^{-3} \mathrm{~Hz}$ to the maximum frequency $f_{\max}=0.5 \mathrm{~Hz}$. This involves dividing the frequencies into $1000$ intervals of equal logarithmic spacing. However, we retain all the original values for frequencies between the minimum frequency $f_{\min}=3 \times 10^{-5} \mathrm{~Hz}$ and $f=10^{-3} \mathrm{~Hz}$. Consequently, the resulting data set comprises a total of $1971$ bins per chunk, effectively balancing computational efficiency while preserving the fidelity of the original data.
We introduce a redefined data set within the coarse-grained frequency region as
\m
f_k &\equiv& \sum_{j \in \operatorname{bin} k} w_j f_j, \\
\bar{D}_k &\equiv& \sum_{j \in \operatorname{bin} k} w_j \bar{D}_j,
\n
where the weights are
\begin{equation}
w_j =\frac{\mathcal{D}^\mathrm{th}(f_j, \vec{\theta}, \vec{n})^{-1}}{\sum_{l \in \mathrm{bin} k}\mathcal{D}^\mathrm{th}(f_l, \vec{\theta}, \vec{n})^{-1}},
\end{equation}
corresponding to the $j$-th bin inside the ``macro-bin" $k$. Note that we have defined $\mathcal{D}^\mathrm{th}(f, \vec{\theta}, \vec{n}) \equiv  \Omega_\mathrm{G W}(f, \vec{\theta})+ \Omega_{\alpha} (f, \vec{n})$ as our model for the data. The noise parameters are denoted as $\vec{n} \equiv \{A, P\}$, while the signal parameters are represented as $\vec{\theta} \equiv \{A_1, \alpha_1, A_2, \alpha_2, A_\mathrm{ECB}, \alpha_\mathrm{ECB}, p, G\mu\}$.  
In practice, we work on the AET basis, and the covariance matrix is diagonal.

The full likelihood is composed of two parts: the Gaussian component and the log-normal component~\cite{Flauger:2020qyi}, namely
\begin{equation}
\ln \mathcal{L}=\frac{1}{3} \ln \mathcal{L}_\mathrm{G}+\frac{2}{3} \ln \mathcal{L}_\mathrm{LN}.
\end{equation}
The Gaussian part is
\begin{equation}
\ln \mathcal{L}_\mathrm{G}(D \mid \vec{\theta}, \vec{n})=-\frac{N_{\mathrm{c}}}{2} \sum_{\alpha} \sum_k n_{\alpha}^{(k)}\left[\frac{\mathcal{D}_{\alpha}^\mathrm{t h}\left(f_{\alpha}^{(k)}, \vec{\theta}, \vec{n}\right)-\mathcal{D}_{\alpha}^{(k)}}{\mathcal{D}_{\alpha}^\mathrm{t h}\left(f_{\alpha}^{(k)}, \vec{\theta}, \vec{n}\right)}\right]^2,
\end{equation}
where the indices $\alpha \in \{\mathrm{A, E, T}\}$ run over the different channel combinations, and the index $k$ runs over the coarse-grained data points.
The log-normal part is 
\begin{equation}
\ln \mathcal{L}_\mathrm{L N}(D \mid \vec{\theta}, \vec{n})=-\frac{N_{\mathrm{c}}}{2} \sum_{\alpha} \sum_k n_{\alpha}^{(k)} \ln ^2\left[\frac{\mathcal{D}_{\alpha}^\mathrm{t h}\left(f_{\alpha}^{(k)}, \vec{\theta}, \vec{n}\right)}{\mathcal{D}_{\alpha}^{(k)}}\right].
\end{equation}

We employ the \texttt{dynesty}~\cite{Speagle:2019ivv} sampler, incorporated within the \texttt{Bilby} package~\cite{Ashton:2018jfp,Romero-Shaw:2020owr}, to explore the parameter space. The model parameters and their corresponding priors are summarized in \Table{tab:priors}. 
In \Fig{posts_Taiji}, we present the posterior distributions for the model parameters inferred by the Taiji detector.
The injected values for each parameter are highlighted in red.
Notably, all model parameters can be accurately recovered within the $2\sigma$ confidence level. 
The recovered values, along with their median and $90\%$ equal-tail uncertainties, are also summarized in \Table{tab:priors}. 
It is observed that the noise parameters, $A$ and $P$, can be measured with a relative uncertainty of $0.4\%$ and $0.03\%$, respectively. 
The parameters associated with the ECB and DWD foregrounds exhibit relatively larger uncertainties due to their weaker signal compared to the SGWB of cosmic strings. 
In contrast, the parameters related to cosmic strings, $\log_{10} p$ and $\log_{10} G\mu$, can be determined with unprecedented accuracy, featuring relative uncertainties of $0.6\%$ and $0.3\%$ respectively.
This indicates a significant improvement of about $70$ and $20$ times in precision compared to the NANOGrav 15-year data set, highlighting the enhanced measurement capability for the model parameters with Taiji. A comparison of the posteriors derived from the NANOGrav 15-year data set and the Taiji detector is presented in \Fig{compare}.

\section{\label{conclusion}Conclusion}

In this paper, we have explored the prospects of detecting an SGWB from cosmic strings using the Taiji space-based GW detector. We assumed that the stochastic signal detected by PTAs, such as NANOGrav, originates from cosmic strings and used the NANOGrav 15-year data set to constrain the model parameters. By analyzing simulated Taiji data and applying parameter estimation techniques, we quantified the uncertainties in determining the model parameters associated with the GW signal from cosmic strings. We demonstrate a significant improvement in precision compared to the NANOGrav 15-year data set, surpassing it by an order of magnitude. This highlights the enhanced measurement capabilities of Taiji. Therefore, space-based GW detectors like Taiji play a crucial role in verifying the stochastic signal detected by PTAs and serve as invaluable tools for further validation of cosmic string signatures. Our study also confirms the importance of accurately characterizing various components of the model~\cite{Boileau:2021gbr}, including the Taiji noise model, and foreground contributions from DWDs and ECBs.

In conclusion, the prospects for Taiji to detect an SGWB from cosmic strings appear promising, and our study provides valuable insights into the parameter estimation and detection capabilities of Taiji for cosmic strings. Future observations and analyses from Taiji and other space-based GW detectors will continue to shed light on the origin and properties of the SGWB, contributing to our understanding of the early Universe and fundamental physics.

\section*{Acknowledgments}
ZCC is supported by the National Natural Science Foundation of China (Grant No. 12247176 and No. 12247112) and the China Postdoctoral Science Foundation Fellowship No. 2022M710429. 
QGH is supported by the grants from NSFC (Grant No.~12250010, 11975019, 11991052, 12047503), Key Research Program of Frontier Sciences, CAS, Grant No.~ZDBS-LY-7009, CAS Project for Young Scientists in Basic Research YSBR-006, the Key Research Program of the Chinese Academy of Sciences (Grant No.~XDPB15). 
CL is supported by the National Natural Science Foundation of China Grant No. 12147132.
LL is supported by the National Natural Science Foundation of China (Grant No. 12247112 and No. 12247176) and the China Postdoctoral Science Foundation Fellowship No. 2023M730300. 
ZY is supported by the National Natural Science Foundation of China under Grant No. 12205015 and the supporting fund for young researcher of Beijing Normal University under Grant No. 28719/310432102.
ZQY is supported by the National Natural Science Foundation of China under Grant No. 12305059.

\bibliographystyle{JHEP}
\bibliography{ref}
\end{document}